\documentclass[9pt,twocolumn,twoside]{osajnl}

\usepackage{amsmath}
\usepackage{amssymb}

\newcommand{\bs}{\boldsymbol}

\journal{ol} % Choose journal (ao, aop, josaa, josab, ol, optica, pr)

\setboolean{shortarticle}{true}
\title{Three-dimensional polarization of fields radiated by partially coherent electromagnetic cylindrical sources}
\author[1]{M. Santarsiero}
\author[2]{J. C. G. De Sande}
\author[3,*]{O. Korotkova}
\author[4]{R. Mart\'{i}nez-Herrero}
\author[4]{G. Piquero}
\author[1]{F. Gori}

\affil[1]{Dipartimento di Ingegneria Industriale, Elettronica e Meccanica, Universit\'a Roma Tre, Via V. Volterra 62, 00146 Rome, Italy}
\affil[2]{ETSIS de Telecomunicación, Universidad Polit\'ecnica de Madrid, Campus Sur 28031 Madrid, Spain}
\affil[3]{Department of Physics, University of Miami, 1320 Campo Sano Drive, Coral Gables, FL, 33146}
\affil[4]{Departamento de \'Optica, Universidad Complutense de Madrid, Ciudad Universitaria, 28040 Madrid, Spain}
\affil[*]{Corresponding author: korotkova@physics.miami.edu
\newline
\newline
© 2024 Optica Publishing Group. One print or electronic copy may be made for personal use only. Systematic reproduction and distribution, duplication of any material in this paper for a fee or for commercial purposes, or modifications of the content of this paper are prohibited.
\newline
\url{https://doi.org/10.1364/OL.486628}
}

\begin{abstract}
Partially coherent electromagnetic sources with cylindrical symmetry and infinite extent radiating outwards are introduced. Their $3\times3$ cross-spectral density matrix is given through expansions of the field components in terms of basis functions related to the Hankel functions. The spectral density and the three-dimensional degree of polarization of such sources and the fields they radiate are examined. 
Several examples are presented and discussed. Among them, a class of cylindrical sources whose coherent vector modes coincide with the above basis functions is defined and studied.
\end{abstract}

\setboolean{displaycopyright}{true}

\begin{document}

\maketitle

Research on partially coherent light sources has been traditionally confined to planar sources \cite{ManWolf95}. However, the interest in non-planar sources, primarily restricted to spherical sources due to the possibility of modeling sunlight radiation \cite{Agarwal:OL04,Divitt:Optica15}, has been growing in recent times to include spatial correlation functions more general with respect to those found in thermal light \cite{Shevchenko:OC05, Gori:OC09, Borghi:OL12}. Our most recent work \cite{deSande:22} has employed spherical harmonics as the modes in the coherent mode representation of partially coherent light, making it possible to model the complete variety of partially coherent spherical sources. Although only planar and spherical geometries have enjoyed the treatment of partially coherent light so far, the Helmholz equation allows for exact solutions with separable variables in eleven coordinate systems \cite{Miller}. This implies that coherent mode representation can be applied to this large variety of geometries for obtaining corresponding extensions to partially coherent cases.

Indeed, in a recent paper of ours~\cite{RMH:OL22} we tackled the problem of defining scalar partially coherent sources on a cylindrical surface and introduced a class of sources of this kind characterized by coherent modes with the form of cylindrical functions. The aim of the present paper is to extend that analysis to three-dimensional (3D) electromagnetic (EM) cases. This will allow us to investigate the polarization characteristics of the field across the source surface and those of the fields they radiate. As we shall see, depending on the cylinder radius and on the distance from the source surface, all three components of the propagated electric field may play a significant role. Thus, this represents a physical example where the concept of 3D polarization is necessary.

Although scalar cylindrical sources were treated in \cite{RMH:OL22}, the same formalism can be used to study  EM sources where the electric field vector ${\bs E}$ is parallel to the cylinder's axis, so that the results presented there remain valid for the EM cylindrical sources in the so-called {\it E-polarization}. In that case, using cylindrical coordinates $r,\varphi,z$ (with $z$ being the cylinder axis) and assuming the field to be independent of $z$, the electric field of spatially coherent radiation on and outside the cylinder was written as a linear combination of outgoing Hankel functions~\cite{Panofsky62,Gbur11}: %, i.~e.,
\begin{equation}
\begin{array}{c}
\displaystyle E_z(r,\varphi) =  \displaystyle \sum_{n=-\infty}^{\infty}  b_{n} \; H_n(kr) \;  e^{{\rm i}  n\varphi}
\;\;\;\;\;\; (r\geq a),  
\end{array}
\label{Ez1}
\end{equation} 
where $k$ is the wavenumber, $\{b_n\}$ a suitable set of coefficients, and $a$ the cylinder radius. Equation~(\ref{Ez1}) is more profitably written as 
\begin{equation}
\begin{array}{c}
\displaystyle E_z(r,\varphi) =  \displaystyle \sum_{n=-\infty}^{\infty}  c_{n} 
\, Z_n(kr) \;  e^{{\rm i} n\varphi}
\;\;\;\;\;\; (r\geq a),  
\end{array}
\label{Ez2}
\end{equation} 
with 
\begin{equation}
c_n = b_n H_n(ka) \; ;
\;\;\;\;\;
Z_n(kr)=\displaystyle\frac{H_n(kr)}{H_n(ka)}
\; .
\label{Ez3}
\end{equation} 
In such a way, 
\begin{equation}
\begin{array}{c}
\displaystyle E_z(a,\varphi) =  \displaystyle \sum_{n=-\infty}^{\infty}  c_{n} 
\;  e^{{\rm i} n\varphi},  
\end{array}
\label{Ez4}
\end{equation} 
and the $c_n$ coefficients correspond to the usual Fourier coefficients of $E_z$ along circle $r=a$.

In order to extend our analysis to the EM domain and study the polarization properties of cylindrical sources, the case of {\it H-polarization} has to be also considered. 
In this case, ${\bs E}$ is replaced by ${\bs B}$ (magnetic induction) and we have
\begin{equation}
\begin{array}{c}
\displaystyle B_z(r,\varphi) =  \displaystyle \sum_{n=-\infty}^{\infty}  d_{n} \; Z_n(kr) \;  e^{{\rm i}  n\varphi}
\;\;\;\;\;\; (r\geq a),  
\end{array}
\label{Bz1}
\end{equation} 
where a different set of coefficients $\{d_n\}$ is used.

On assuming that the field propagates in vacuo, the corresponding components of the electric field turn out to be~\cite{Gbur11, Arfken05}
\begin{equation}
\left\{
\begin{array}{l}
E_{r}(r,\varphi)
=
\displaystyle\frac{{\rm i}c}{kr}
\, \displaystyle\frac{\partial B_z}{\partial \varphi}
=
- c
\displaystyle \sum_{n=-\infty}^{\infty}  n \, d_{n} 
\, \displaystyle\frac{Z_n(kr)}{kr}
 \;  e^{{\rm i}  n\varphi}
 \; ,
\\
%\\
E_{\varphi}(r,\varphi)
=
- \displaystyle\frac{{\rm i} c}{k}
\, \displaystyle\frac{\partial B_z}{\partial r} 
=
- {\rm i}  c
\displaystyle \sum_{n=-\infty}^{\infty}  d_{n} \; Z'_n(kr) \;  e^{{\rm i} n\varphi}
\; ,
 \end{array}
 \right.
\label{Ep1}
\end{equation} 
where $c$ is the speed of light and prime denotes the derivative.

As for the case of E-polarization, we can set $a_n = - {\rm i}  \, c \, d_n$
%
%\begin{equation}
%a_n = - {\rm i}  \, c \, d_n \; ,
%\label{Ep2}
%\end{equation} 
%
and
\begin{equation}
R_n(kr)= 
\displaystyle\frac{- {\rm i} n}{H_n(ka)}
\,
\displaystyle\frac{H_n(kr)}{kr}
\, ,
\;\;\;\;\;
\Phi_n(kr)=\displaystyle\frac{H'_n(kr)}{H_n(ka)} \; ,
\label{Ep3}
\end{equation} 
and obtain, together with Eq.~(\ref{Ez2}), the following expressions for the three components  (namely, radial, tangential, and axial) of the electric field all over the space outside the cylinder:
\begin{equation}
\left\{
\begin{array}{l}
E_{r}(r,\varphi)
=
\displaystyle \sum_{n=-\infty}^{\infty}
a_{n} 
\;
R_n(kr)
 \;  e^{{\rm i}  n\varphi}
 \; ,
\\
E_{\varphi}(r,\varphi)
=
\displaystyle \sum_{n=-\infty}^{\infty}  a_{n} 
\; \Phi_n(kr) \;  e^{{\rm i} n\varphi}
\; ,
\\
\displaystyle E_z(r,\varphi) =  \displaystyle \sum_{n=-\infty}^{\infty}  c_{n} 
\, Z_n(kr) \;  e^{{\rm i} n\varphi}
 \; .
 \end{array}
 \right.
\label{Ep4}
\end{equation} 
Differently from the $c_n$ coefficients, $a_n$ do not represent the Fourier coefficients of any of the two transverse field components along the circle $r = a$ because neither $R_n(ka)$ nor $\Phi_n(ka)$ equal one.
We shall loosely refer to functions $R_n$, $\Phi_n$, and $ Z_n$ as the radial, tangential, and axial basis functions of the field, respectively. 
 
\begin{figure}[!ht]
	\centering
	\includegraphics[width=0.9\linewidth] {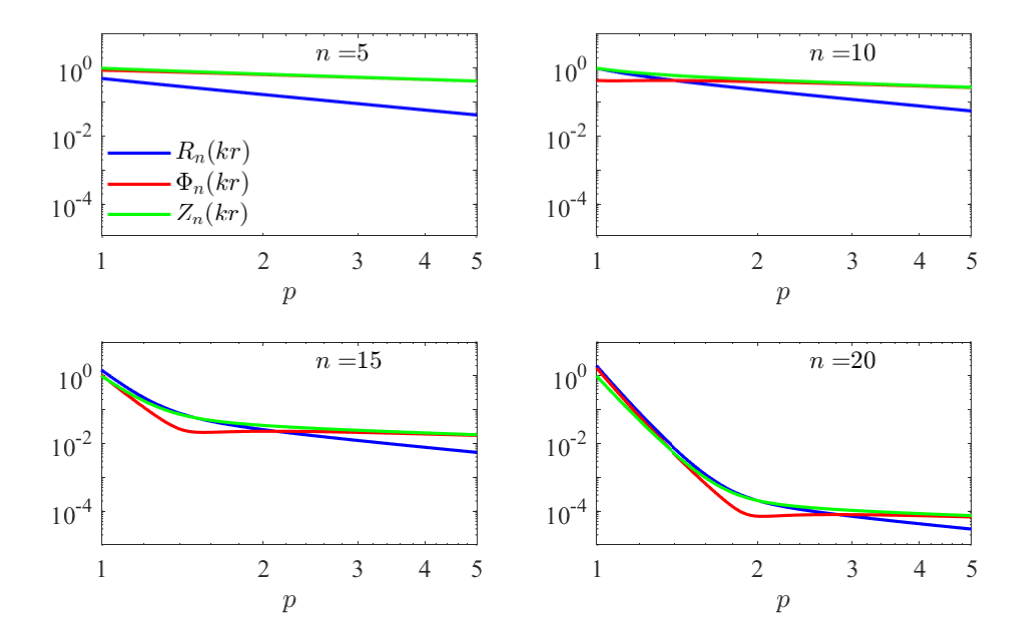} 
\caption{Absolute value of the radial (blue), tangential (red), and axial basis functions (green) \emph{vs} the radial distance ($p=r/a$), for $ka=10$ and different values of order $n$.}
\label{fig1}
\end{figure}

Figure~\ref{fig1} shows the absolute value of the basis functions for $ka=10$ and different orders, as functions of the propagation distance normalized to the cylinder radius ($p=r/a$). 
%%%%%%%%%%%%%%%%%%%%%%%%%%%%%%%%%%%%%%

Since for any value of index $n$ the expansion coefficients of the radial and the tangential components coincide (see Eq.~(\ref{Ep4})), the comparison of the corresponding curves (blue and red, respectively) give also information about the relative weight of such components. 
For $n<ka$ a decrease as $r^{-1/2}$ can be recognized for both the tangential and the axial components (red and green curves, almost coinciding), while a faster decrease, as $r^{-3/2}$, results for the radial component (blue curve), which becomes less and less significant on moving away from the cylinder surface. 
Such behaviors can be deduced analytically from the asymptotic expressions of the Hankel functions~\cite{Erdelyi53}. 
When $n$ approaches $ka$ the radial component becomes more significant, and even dominant with respect to the tangential one, but only near the cylinder surface. 
When $n$ exceeds $ka$ all the components decrease very quickly with the distance from the surface, with a rate that increases on increasing $n$ (approximately as $r^{-18}$ for $n=20$), and become of the order $10^{-3}$ the initial ones already at  $r/a\approx 2$. 
As we shall see, the above behaviors fully determine the evolution of the field components upon propagation. 

%%%%%%%%%%%%%%%%%
 %

%
%

As a matter of fact, the effect of propagation, starting from the cylinder surface, can be traced back to spatial filtering of the expansion coefficients, as it can be easily recognized if we write Eq.~(\ref{Ep4}) in the following form:
\begin{equation}
\left\{
\begin{array}{l}
E_{r}(r,\varphi)
=
\displaystyle \sum_{n=-\infty}^{\infty}
a_{n} 
\left[
\displaystyle\frac{R_n(kr)}{R_n(ka)}
\right]
R_n(ka)
 \;  e^{{\rm i}  n\varphi}
 \; ,
\\
E_{\varphi}(r,\varphi)
=
\displaystyle \sum_{n=-\infty}^{\infty}  a_{n} 
\left[
\displaystyle\frac{\Phi_n(kr)}{\Phi_n(ka)}
\right]
\Phi_n(ka) 
\;  e^{{\rm i} n\varphi}
\; ,
\\
\displaystyle E_z(r,\varphi) =  \displaystyle \sum_{n=-\infty}^{\infty}  c_{n} 
\left[
\displaystyle\frac{Z_n(kr)}{Z_n(ka)}
\right]
Z_n(ka) 
 \;  e^{{\rm i} n\varphi}
 \; .
 \end{array}
 \right.
\label{Ep4bis}
\end{equation} 
%

%\begin{equation}
%\rho_n(r)
%=
%\displaystyle\frac{R_n(kr)}{R_n(ka)}
% \; ,
% \;\;
%\phi_n(r)
%=
%\displaystyle\frac{\Phi_n(kr)}{\Phi_n(ka)}
% \; ,
% \;\;
%\zeta_n(r)
%=
%\displaystyle\frac{Z_n(kr)}{Z_n(ka)}
% \; .
%\label{Ep4ter}
%\end{equation}

\begin{figure}[!ht]
	\centering
	\includegraphics[width=0.95\linewidth]{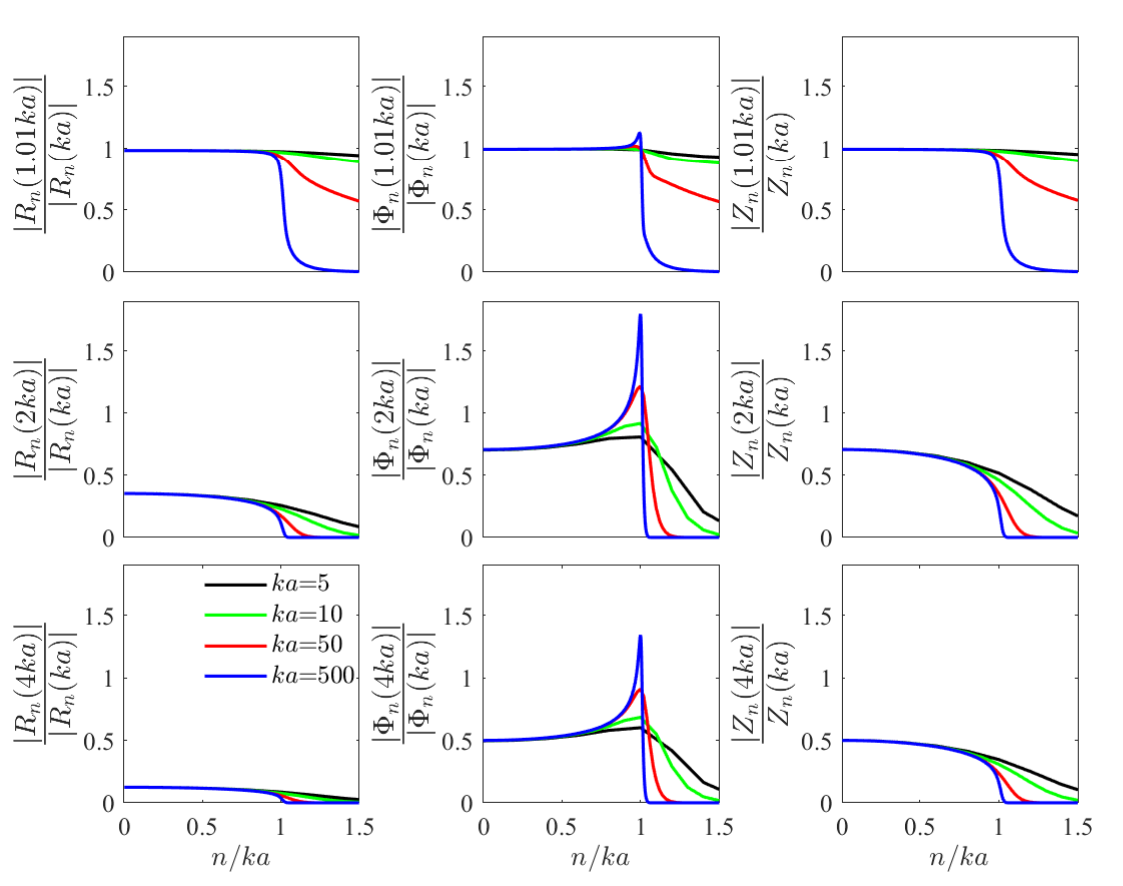} 
\caption{Absolute value of the weighting functions for different values of the radial distance ($p=r/a$) and the cylinder radius ($ka$), as functions of order $n$ normalized to $ka$.}
\label{fig2}
\end{figure}

The quantities in square brackets represent the weights by which the coefficients are multiplied during propagation. Their absolute values are shown in Fig.~\ref{fig2} as functions of $n/ka$, for different choices of the cylinder radius $a$ and the propagation distance $p=r/a$. Although $n$ takes integer values, continuous curves are presented for better visualization.

Some comments are worthwhile. First, the radial component becomes negligible with respect to the other ones, for any $n$, when the propagation distance exceeds some multiple of the cylinder radius. Second, all the components vanish for values of $n$ sufficiently greater than $ka$, except very close to the surface. Since the $ka$ parameter expresses the ratio between the circumference of the cylinder base and the wavelength, and the expanding basis functions are accompanied by the angular Fourier component $\exp({\rm i} n \varphi)$, when $n$ exceeds $ka$ the involved phase details become smaller than the wavelength. The cutoff around $n=ka$ means that such contributions do not propagate, exactly as it happens for the evanescent waves produced by planar sources. Such a behavior was already observed in Ref.~\cite{RMH:OL22} for the axial component.
Third, an enhancement in the propagating tangential component with respect to that across the cylinder is observed for values of $n$ close to $ka$.

Let us now pass to the partially coherent case, where the $a_n$ and $c_n$  coefficients become random variables. We make use of the $3 \times 3$ cross-spectral density matrix $\widehat W({\bs r}_1,{\bs r}_2)$~\cite{ManWolf95}, which gives account of the two-point, second-order correlations among all the field components and whose elements are defined as
\begin{equation}
W_{st}({\bs r}_1,{\bs r}_2)
=
\langle
E_s^*({\bs r}_1)
E_t({\bs r}_2)
\rangle
\;\;\;\;
(s,t=r, \varphi, z)
\; ,
\label{CSD}
\end{equation} 
the angle brackets denoting ensemble average and the star standing for complex conjugate. In particular, the local properties of the field at point ${\bs r}$ are taken into account by the matrix
\begin{equation}
\widehat P({\bs r})
=
\widehat W({\bs r},{\bs r})
\; ,
\label{polmat}
\end{equation} 
referred to as the polarization matrix. The spectral density of the field is
\begin{equation}
S({\bf r})={\rm Tr}\{ \widehat P({\bs r})\}
\; ,
\label{spectral}
\end{equation} 
with Tr$\{\cdot\}$ denoting the trace, and a 3D degree of polarization (DOP) can be defined as~\cite{Setala:PRE02} 
\begin{equation}
P_{3}({\bs r})
=
\sqrt{
\displaystyle\frac{3}{2}
\left[
\displaystyle\frac{{\rm Tr}\{\widehat P^2({\bs r})\}}{{\rm Tr}^2\{\widehat P({\bs r})\}}
-
\displaystyle\frac{1}{3}
\right]
}
\; ,
\label{poldeg}
\end{equation} 
which is always limited to the interval $[0,1]$. 
According to the above definition, the DOP of a completely unpolarized 2D field is 1/2, denoting that a certain ``polarization'' exists, due to the fact that one of the three field components is missing. It should be recalled that alternative definitions have been proposed for a 3D DOP that give 0 for a completely unpolarized 2D field~\cite{Ellis:OC05,Aunon:OL13}.

According to Eqs.~(\ref{Ep4}), (\ref{CSD}), (\ref{polmat}), and (\ref{poldeg}), the polarization properties of any field endowed with cylindrical symmetry are determined by  parameters 
$\langle a^*_n a_m \rangle$, $\langle c^*_n c_m \rangle$, and $\langle a^*_n c_m \rangle$, which give the correlations among all the modes involved in the field expansions. The presence of three orthogonal, partially correlated field components reflects on the DOP of the field across the source and after propagation.
We can give now some examples where particular choices of the correlations among the modes give rise to different behaviors of the DOP across the cylinder source and upon propagation. 

One of the most significant cases appears if all the involved coefficients are set to be mutually uncorrelated, that is,
\begin{equation}
	\langle
	a^*_n
	a_m
	\rangle
	=
	\alpha_n
	\delta_{nm}
	\; ; \;\;\;\; 
	\langle
	c^*_n
	c_m
	\rangle
	=
	\gamma_n
	\delta_{nm}
	\; ; \;\;\;\; 
	\langle
	c^*_n
	a_m
	\rangle
	=0
	\; ,
	\label{coeffEs4}
\end{equation} 
where $\alpha_n$ and  $\gamma_n$ are positive quantities and $\delta_{nm}$ is the Kronecker delta.
This means that the fields $\left[ R_n({\bs r}) \hat{r} + \Phi_n({\bs r}) \hat{\varphi}\right] \exp{({\rm i} n\varphi)}$ and $Z_n({\bs r}) \exp{({\rm i} n\varphi)} \hat{z}$ are proportional to the coherent vector modes of the partially coherent cylindrical source, with eigenvalues (proportional to) $\alpha_n$ and $\gamma_n$, respectively~\cite{Gori:JOSAA03}. With such a choice both the spectral density and the DOP turn out to be independent of $\varphi$ for any set of eigenvalues. In fact, the elements of $\hat P$ reduce to
\begin{equation}
\begin{array}{l}
P_{r r}({\bs r})
=
\displaystyle \sum_{n}  
\alpha_n
|R_n(kr)|^2
\; ,
\\
P_{\varphi \varphi}({\bs r})
=
\displaystyle \sum_{n}  
\alpha_n
|\Phi_n(kr)|^2
\; ,
\\
P_{zz}({\bs r})
=
\displaystyle \sum_{n}  
\gamma_n
|Z_n(kr)|^2
\; ,
\\
P_{r \varphi}({\bs r})
=
P^*_{\varphi r}({\bs r})
=
\displaystyle \sum_{n}  
\alpha_n
R^*_n(kr) 
\Phi_n(kr) 
\; ,
\\
P_{r z}({\bs r})
=
P^*_{zr}({\bs r})
=
P_{\varphi z}({\bs r})
=
P^*_{z \varphi}({\bs r})
=
0
\; .
\end{array}
\label{Pelem}
\end{equation} 

\begin{figure}[!ht]
	\centering
	\includegraphics[width=0.9\linewidth]{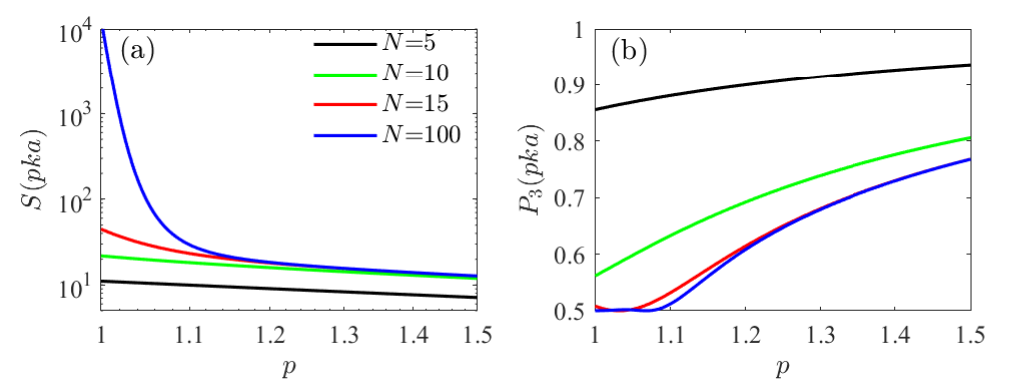}
	\caption{(a) Spectral density and (b) 3D DOP outside the cylindrical source, for H polarization, from Eq.~(\ref{coeff3Es1}), with $ka=10$ and several values of $N$.}
	\label{fig3}
\end{figure}

As a first example of this case, we consider a field for which only the H-polarization is present, i.e., $\gamma_n=0$, $\forall n$.  This is the complementary problem to the one studied in \cite{RMH:OL22}, where a detailed analysis of the E-polarization was performed. Furthermore,  we set the $\alpha$ coefficients as: 
\begin{equation}
\alpha_{n} = A \;\;\; (|n|\le N) ;
\hspace{.5cm}
\alpha_{n} = 0 \;\;\; (|n|> N),
\label{coeff3Es1}
\end{equation} 
where $A>0$. The effect of changing the value of $N$ is shown in Fig.~\ref{fig3}, where $S(r)$ and $P_{3}(r)$ are plotted as  functions of $p$ for $ka = 10$ and different values of $N$. It is seen that the propagating spectral density (Fig.~\ref{fig3}a) increases with $N$ but, except at short distances from the cylinder surface, only up to a value $N\approx ka$. Such a behavior, analogous to the one observed for E-polarization \cite{RMH:OL22} denotes that modes with $N>ka$ do not propagate and remain confined near the cylinder surface. On the other hand (Fig.~\ref{fig3}b) $P_3$ always tends to 1 when $r \to \infty$, and this is due to the fact that the radial component of the field becomes more and more negligible on increasing $r$ and perfect polarization in reached in this limit. Near the cylinder, if $N$ is large enough, $P_3$  may also take values close to 1/2 at some distance from the surface, indicating that there the two field components are almost uncorrelated and carry the same power.

\begin{figure}[!ht]
	\centering
        \includegraphics[width=0.9\linewidth]{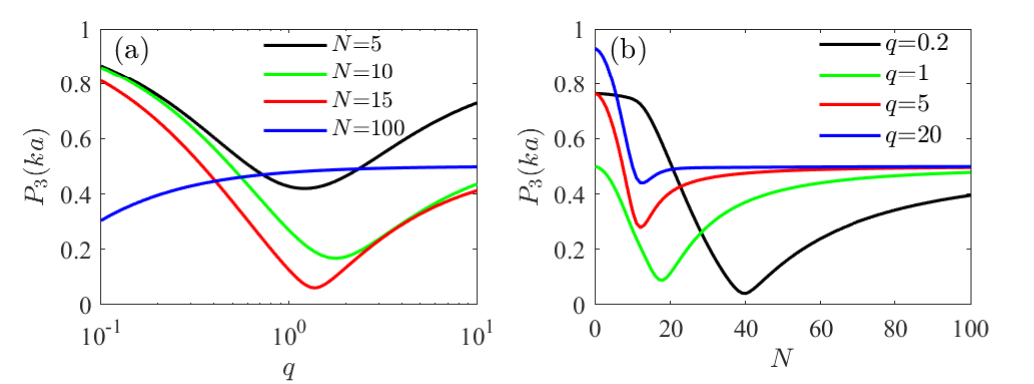} \\
	\includegraphics[width=0.9\linewidth]{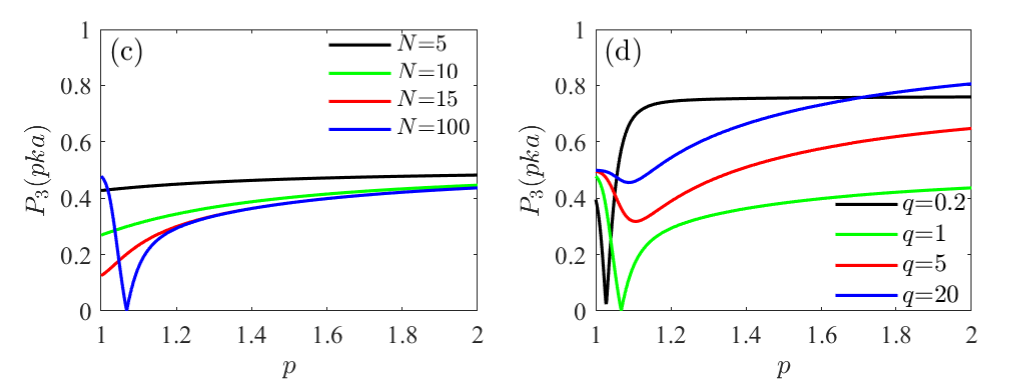}
	\caption{3D DOP for a cylindrical source, from Eq.~(\ref{coeff3Es4b}) with $ka=10$ and different choices of $N$ and $q$. (a)-(b) on the source surface: (a) as a function of $q$; (b) as a function of $N$. (c)-(d) as a function of $p$:
 (c) for $q=1$; (d) for $N=100$. }
	\label{fig6}
\end{figure}

Adding the modes polarized along $z$ significantly changes the behavior of the DOP during propagation. In the following example, we take
\begin{equation}
	\alpha_{n} = A; \; \gamma_{n} = C \; (|n|\le N) ;
	\hspace{.5cm}
	\alpha_{n} = \gamma_{n} = 0  \; (|n|> N)
	\label{coeff3Es4b}
\end{equation} 
and the corresponding behavior of the DOP as a function of $p$ is shown in Fig.~\ref{fig6} (a)-(b) (on the surface) and Fig.~\ref{fig6}(c)-(d) (upon propagation) for $ka=10$. 
%\textcolor{red}{(a) for different values of $N$ (and fixed ratio $q=A/C$) and (b) for different values of the ratio $q$ (and fixed $N$).}
It can be noted that the DOP on the source surface varies from zero to nearly 1. This means that, by appropriately selecting the source parameters, almost any DOP can be obtained [see Fig.~\ref{fig6} (a)-(b)]. On the other hand, when $A=C$ and $N$ is high enough, the field is completely depolarized at a certain distance from the surface [Fig.~\ref{fig6} (c)]. After this distance, the DOP grows towards a maximum value, which, for fixed $N$, depends on the ratio $q=A/C$ [Fig.~\ref{fig6} (d)].
%
%\begin{figure}[!ht]
%	\centering
%	\includegraphics[width=0.9\linewidth]{Figure4}
%	\caption{3D  DOP on the source from Eq.~(\ref{coeff3Es4b}) with $ka=10$; (a) as a function of $q=A/C$; (b) as a function of $N$.}
%	\label{fig7}
%\end{figure}
%
%
%\begin{figure}[!ht]
%	\centering
%	\includegraphics[width=0.9\linewidth]{Figure5}
%	\caption{3D DOP outside a cylindrical source, from Eq.~(\ref{coeff3Es4b}), with $ka=10$; (a) $q=A/C=1$ and some values of $N$; (b) $N=100$ and some values of $q$. }
%	\label{fig6}
%\end{figure}
%
%

\begin{figure}[ht]
	\centering
	\includegraphics[width=0.9\linewidth]{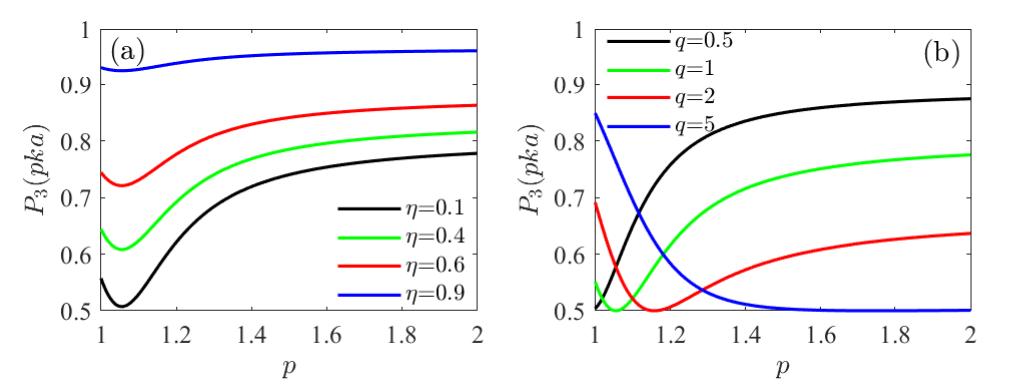}
	\caption{3D DOP outside a cylindrical source, from Eq.~(\ref{coeffEs3}), with $ka=10$, $n_1=12$, $n_2=8$; (a) $q=A/C=1$ and several values of $\eta$; (b) $\eta=0$ and several values of the ratio $q=A/C$.}
	\label{fig5}
\end{figure}

More varied behaviors can be obtained if some correlations are allowed among the coefficients, as will be shown in the following two examples (please note that in this way the basis functions do not represent the coherent modes of the source). In the first case, we introduce a correlation between the two polarizations. For simplicity, we suppose that only one mode contributes to each of the two polarizations, say, the order $n_1$ for H polarization and the order $n_2$ for E polarization. This means that the correlations take the form
\begin{equation}
\begin{array}{c}
\langle
a^*_{n}
a_{m}
\rangle
=
A \, \delta_{n n_1}\delta_{m n_1}
;
\;\;\;
\langle
c^*_{n}
c_{m}
\rangle
=
C \, \delta_{n n_2}\delta_{m n_2}
;
\\
\langle
a^*_{n}
c_{m}
\rangle
=
\eta \, \sqrt{A \, C} \; \delta_{n n_1}\delta_{m n_2}
,
\end{array}
\label{coeffEs3}
\end{equation} 
where $A$ and $C$ are positive quantities and $\eta$ is a complex number giving account of the correlation between the modes and whose absolute value ranges from 0 and 1. In such a case the 3D DOP can be evaluated from Eqs.~(\ref{Ep4})-(\ref{poldeg}) as
\begin{equation}
P_{3}({\bs r})
=
\sqrt{
1 
-
\displaystyle\frac{3 \, S_z(r) S_\perp(r)} {[S_z(r) + S_\perp(r)]^2} \; (1-|\eta|^2)
}
\; ,
\label{poldeg_Es3}
\end{equation} 
where for brevity the spectral densities associated with the field components across the transverse plane and along the $z$ axis have been introduced, namely, $S_\perp(r)= A [|R_{n_1}({\bs r})|^2+|\Phi_{n_1}({\bs r})|^2]$ and $S_z(r)= C |Z_{n_2}({\bs r})|^2$. It is seen that even in this case such quantities depend only on the radial distance $r$, and the same happens for the DOP. Figure~\ref{fig5} shows that the 3D DOP never goes below 1/2. This is due to the fact that the electric field is obtained as the superposition of two totally polarized components and behaves as a 2D field, as far as its polarization is concerned.

\begin{figure}[ht]
	\centering
	\includegraphics[width=0.8\linewidth]{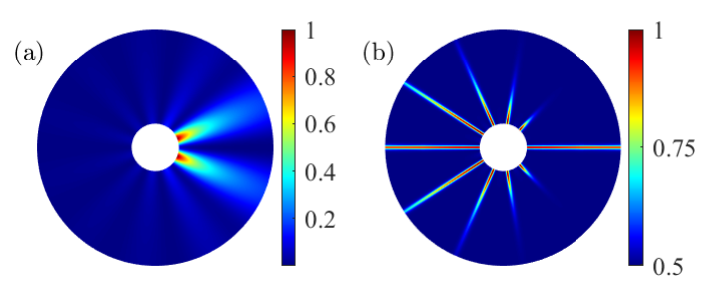}
	\caption{ (a) Spectral density, divided by its maximum, and (b) 3D  DOP outside a cylindrical source, from Eq.~(\ref{coeffEs2}), with $ka=10$ and $N=5$. The range of the radial variable is $a \le r \le 5a$.}
 %maximum propagation distance is $5ka$.}
	\label{fig4}
\end{figure}

In the last example the two polarizations are completely mutually uncorrelated, but the coefficients for each polarization are supposed to present a non-zero correlation. In particular, we set
\begin{equation}
\langle
a^*_{n}
a_{m}
\rangle
=
\langle
c^*_{n}
c_{m}
\rangle
= n m
;
\;\;\;
\langle
a^*_{n}
c_{m}
\rangle
= 0
\;\;\;
(|n|,|m| \le N)
\; ,
\label{coeffEs2}
\end{equation} 
and all coefficients are supposed to vanish if $|n|> N$ or $|m|> N$. 
Figure~\ref{fig4} shows the corresponding $S(r)$ and $P_{3}(r)$ in false colors across the transverse plane, for $ka = 10$ and different values of $N$.
Contrary to the previous cases, here there is a clear angular dependence of the spectral density and the 3D DOP. In particular, two directions can be identified where the spectral density is very large. The 3D DOP is practically 0.5 in the whole space, except along some directions where it increases to 1. These directions coincide with those in which either the E or the H-polarization cancels out.

In summary, the new class of partially coherent EM radiation originating from the surface of an infinite cylinder and propagating outwards is introduced, using the coherent mode representation separable in cylindrical geometry. Due to the qualitatively different behavior of the three components of the electric field, the evolution of the spectral density and the degree of polarization (being intrinsically three-dimensional in this case) is shown to be rich in possible distributions while fully accountable. 

	\begin{backmatter}
		\bmsection{Funding} 
		Spanish Ministerio de Econom\'ia y Competitividad, project PID2019-104268 GB-C21. OK acknowledges the Copper Fellowship program at the University of Miami.
	
		\bmsection{Disclosures} The authors declare no conflicts of interest.
		
		\bmsection{Data availability} No data were generated or analyzed in the presented research.
		
		%\bmsection{Supplemental document} See Supplement 1 for supporting content. 
		
	\end{backmatter}

\end{document}